\documentclass[oneside,english]{book}
\usepackage{lmodern}
\usepackage{lmodern}
\usepackage[T1]{fontenc}
\usepackage[latin9]{inputenc}
\usepackage{geometry}
\usepackage{float}
\usepackage{textcomp}
\usepackage{amsmath}
\usepackage{graphicx}
\PassOptionsToPackage{normalem}{ulem}
\usepackage{ulem}

\makeatletter

\providecommand{\tabularnewline}{\\}

\usepackage{datetime}
\newdateformat{monthyeardate}{%
  \monthname[\THEMONTH], \THEYEAR}

\usepackage{babel}

\usepackage{babel}

\usepackage{babel}

\makeatother

\usepackage{babel}
\begin{document}
\thispagestyle{empty}

\vspace*{3cm}

\begin{center}
\textbf{\huge{}The JETSPIN User Manual} 
\par\end{center}

\bigskip{}

\begin{center}
{\large{}M. Lauricella$^{*}$, G. Pontrelli$^{*}$, I. Coluzza$^{\#}$,
D. Pisignano$^{\text{\textdegree}}$, S. Succi$^{*}$}
\par\end{center}{\large \par}

\medskip{}

\begin{center}
{\large{}$^{*}$Istituto per le Applicazioni del Calcolo CNR,}
\par\end{center}{\large \par}

\begin{center}
{\large{}Via dei Taurini 19, 00185 Rome, Italy} 
\par\end{center}

\smallskip{}

\begin{center}
{\large{}$^{\#}$Faculty of Physics, University of Vienna,}
\par\end{center}{\large \par}

\begin{center}
{\large{}Boltzmanngasse 5, 1090 Vienna, Austria}
\par\end{center}{\large \par}

\smallskip{}

\begin{center}
{\large{}$^{\text{\textdegree}}$Dipartimento di Matematica e Fisica
``Ennio De Giorgi'', University of Salento,}
\par\end{center}{\large \par}

\begin{center}
{\large{}via Arnesano, 73100 Lecce, Italy}
\par\end{center}{\large \par}

\medskip{}

\begin{center}
\textbf{\large{}version 1.20 - \monthyeardate\today}
\par\end{center}{\large \par}

\newpage{}

\setcounter{page}{1}

\newpage{}

\vspace*{3cm}

\begin{center}
\textbf{\Large{}Disclaimer}
\par\end{center}{\Large \par}

\bigskip{}

None of the authors, nor any contributor to the JETSPIN code or its
derivatives guarantee that the software and associated documentation
is free from error. Neither do they accept responsibility for any
loss or damage that results from its use. The responsibility for ensuring
that the software is fit for purpose lies entirely with the user.

\newpage{}

\vspace*{3cm}

\begin{center}
\textbf{\Large{}JETSPIN Acknowledgements} 
\par\end{center}

\bigskip{}

\begin{center}
The software development process has received funding from the European
Research Council under the European Union's Seventh Framework Programme
(FP/2007-2013)/ERC Grant Agreement n. 306357 ``NANO-JETS''. 
\par\end{center}

\newpage{}

\vspace*{3cm}

\begin{center}
\textbf{\Large{}About JETSPIN}
\par\end{center}{\Large \par}

\bigskip{}

The code is licensed under Open Software License v. 3.0 (OSL-3.0).
The full text of the licence can be found on the website: http://opensource.org/licenses/OSL-3.0

If results obtained with this code are published, an appropriate citation
would be:

Marco Lauricella, Giuseppe Pontrelli, Ivan Coluzza, Dario Pisignano,
Sauro Succi, JETSPIN: A specific-purpose open-source software for
electrospinning simulations of nanofibers, \textit{Submitted to Computer
Physics Communications}, 2015.

\newpage{} \tableofcontents{}

\newpage{}

\chapter{Introduction\label{chap:Introduction}}

\section*{Scope of Chapter}

This chapter describes the design and directory structure of JETSPIN.

\section{The JETSPIN Package\label{sec:The-JETSPIN-Package}}

In the recent years, electrospun nanofibers have gained a considerable
industrial interest due to many possible applications, such as tissue
engineering, air and water filtration, drug delivery and regenerative
medicine. In particular, the high surface-area ratio of the fibers
offers an intriguing prospect for technological applications. As consequence,
several studies were focused on the characterization and production
of uni-dimensionally elongated nanostructures. A number of reviews
\cite{li2004electrospinning,greiner2007electrospinning,carroll2008nanofibers,huang2003review,persano2013industrial}
and books \cite{yarin2014fundamentals,wendorff2012electrospinning,pisignanoelectrospinning}
concerning electrospinning have been published in the last decade.

Typically, electrospun nanofibers are produced at laboratory scale
by the uniaxial stretching of a jet, which is ejected at the nozzle
from a charged polymer solution. The initial elongation of a jet can
be produced by applying an externally electrostatic field between
the spinneret and a conductive collector. Electrospinning involves
mainly two sequential stages in the uniaxial elongation of the extruded
polymer jet: an initial quasi steady stage, in which the electric
field stretches the jet in a straight path away from the nozzle of
the ejecting apparatus, and a second stage characterized by a bending
instability induced from small perturbations, which misalign the jet
from its axis of elongation \cite{zeng2006numerical}. These small
disturbances may originate from mechanical vibrations at the nozzle
or from hydrodynamic-aerodynamic perturbations within the experimental
apparatus. Such a misalignment provides an electrostatic-driven bending
instability before the jet reaches the conductive collector, where
the fibers are finally deposited. As a consequence, the jet path length
between the nozzle and the collector increases, and the stream cross-section
undergoes a further decrease. The ultimate goal of electrospinning
process is to minimize the radius of the collected fibers. By a simple
argument of mass conservation, this is tantamount to maximizing the
jet length by the time it reaches the collecting plane. By the same
argument, it is therefore of interest to minimize the length of the
initial stable jet region. Consequently, the bending instability is
a desirable effect, as it produces a higher surface-area-to-volume
ratio of the jet, which is transferred to the resulting nanofibers
\cite{feng2003stretching}.

Computational models provide a useful tool to elucidate the physical
phenomenon and provide information which might be used for the design
of electrospinning experiments. Numerical simulations can be used
to improve the capability of predicting the key processing parameters
and exert a better control on the resulting nanofiber structure. In
recent years, with a renewed interest in nanotechnology, electrospinning
studies attracted the attention of many resarchers both from modeling,
computational and experimental point of view \cite{reneker2000bending,yarin2001taylor,fridrikh2003controlling,theron2004experimental,lu2006computer,zeng2006numerical}.
Although some author uses dissipative particle dynamics (DPD) mesoscale
simulation method into electrospinning study \cite{wang2013simulation},
most models usually treat the jet filament as a series of discrete
elements ({\em beads}) obeying the equations of continuum mechanics
\cite{reneker2000bending,yarin2001taylor}. Each bead is subject to
different types of interactions, such as long-range Coulomb repulsion,
viscoelastic drag, the external electric field. The main aim of such
models is to explain the complexity of the resulting dynamics and
provide the set of parameters driving the optimal process. The effect
of fast-oscillating loads on the bending instability have been explored
in an extensive modeling and computational study \cite{coluzza2014ultrathin}.

JETSPIN delivers a FORTRAN code especially designed to simulate the
electrospinning process in a variety of different conditions and experimental
settings. This comprehensive platform is devised such that different
cases and input variables can be described and simulated. The framework
is developed to exploit several computational architectures, both
serial and parallel.

JETSPIN, as open-source software, can be used to carry out a systematic
sensitivity analysis over a broad range of parameter values. The results
of simulations provide valuable insight on the physics of the process
and can be used to assess experimental procedures for an optimal design
of the equipment and to control processing strategies of technological
advanced nanofibers.

\section{Structure\label{sec:Structure}}

JETSPIN is written in free format FORTRAN90, and it consists of approximately
160 subroutines. The source exploits the modular approach provided
by the programming language. All the variables having in common description
of certain features or method are grouped in modules. The convention
of explicit type declaration is adopted, and all the arguments passed
in calling sequences of functions or subroutines have defined intent.
We use the PRIVATE and PUBLIC accessibility attributes in order to
decrease error-proneness in programming.

\begin{figure}
\begin{centering}
\includegraphics[scale=0.18]{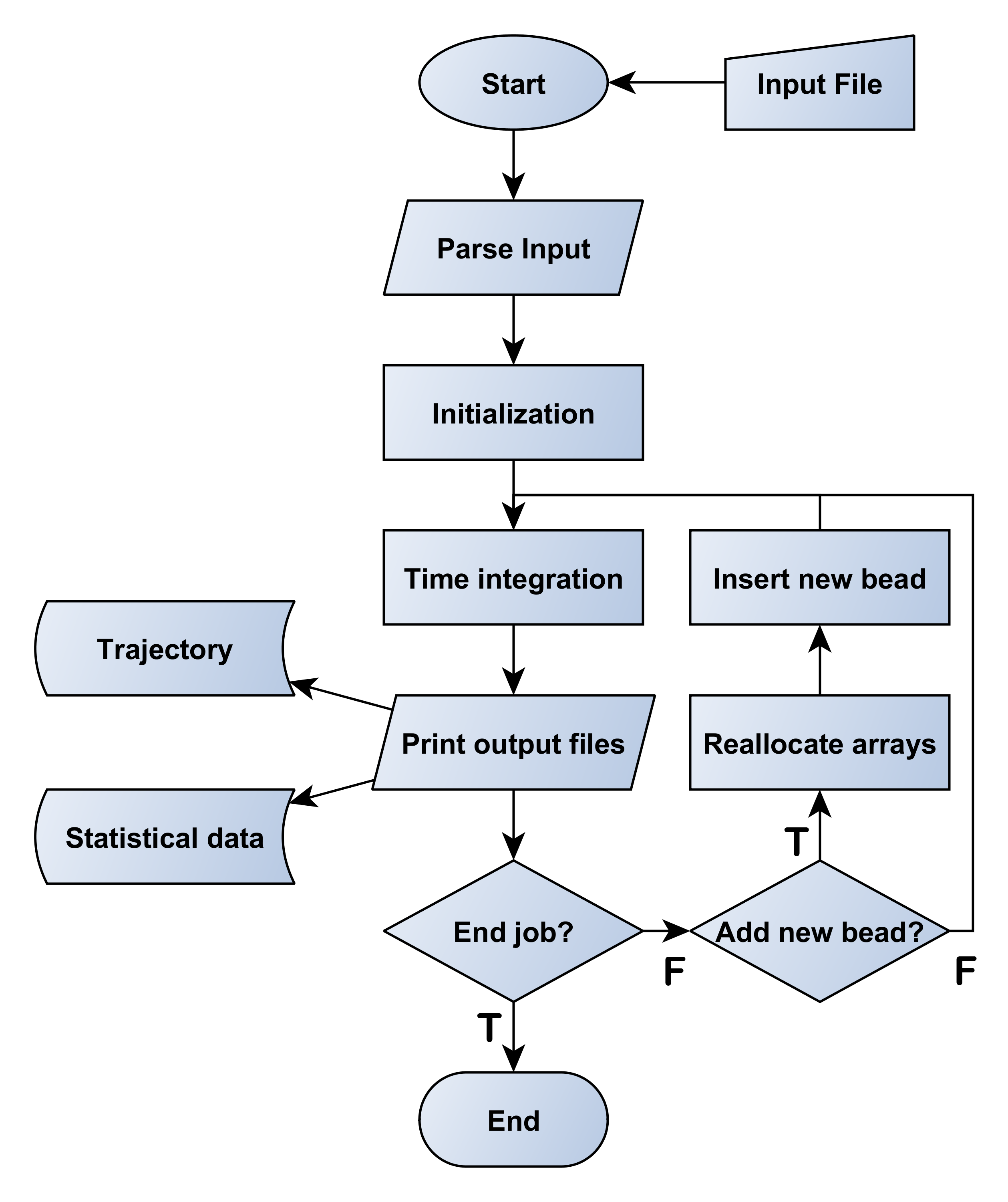} 
\par\end{centering}

\protect\caption{Structure of the main JETSPIN program.}

\label{Fig:Structure-program} 
\end{figure}

The main routines have been gathered in the \texttt{main.f90} file,
which drives all the CPU-intensive computations needed for the capabilities
mentioned below. The variables describing the main features of nanofibers
(position, velocity, etc.) are declared in the \texttt{nanojet\_mod.f90}
file, which also contains the main subroutines for the memory management
of the fundamental data of the simulated system. Since the size system
is strictly time-dependent, JETSPIN exploits the dynamic array allocation
features of FORTRAN90 to assign the necessary array dimensions. In
particular, the size system is modified by the routines \texttt{add\_jetbead}
and \texttt{erase\_jetbead}, while the decision of the main array
size, declared as \texttt{mxnpjet}, is handled by the routine \texttt{reallocate\_jet}.
The sizes of various service bookkeeping arrays are handled within
a parallel implementation strategy, witch exploits few dedicated subroutines
(see Sec \ref{sec:Parallelization}). All the implemented time integrators
are written in the \texttt{integrator\_mod.f90} file, which contains
the routine \texttt{driver\_integrator} to select the proper integrator,
as indicated in the input file. All the terms of equations of motion
for the implemented model (see Sec \ref{sec:Equation-of-Motion})
are computed by routines located in the \texttt{eom\_mod.f90} file,
which call other subroutines in the files \texttt{coulomb\_force\_mod.f90},
\texttt{viscoelastic\_force\_mod.f90} and \texttt{support\_functions\_mod.f90}.
A summarizing scheme of the main JETSPIN program in the \texttt{main.f90}
file has been sketched in Fig \ref{Fig:Structure-program}.

The user can carry out simulations of nanofibers without a detailed
understanding of the structure of JETSPIN code. All the parameters
governing the system can be defined in the input file (see Sec \ref{sec:Input-file}),
which is read by routines located in the \texttt{io\_mod.f90} and
\texttt{parse\_mod.f90} files. Instead, the user should be acquainted
with the model described in Cap \ref{chap:JETSPIN-Model-and}. The
content of the output file is completely customizable by the input
file as described in Sec \ref{sec:Input-file}, and it can report
different time-averaged observables computed by routines of the module
\texttt{statistic\_mod} (see Sec \ref{sec:Output-files}). The routines
in the \texttt{error\_mod.f90} file can display various warning or
error banners on computer terminal, so that the user can easily correct
the most common mistakes in the input file.

JETSPIN is supplied as a single UNIX compressed (tarred and gzipped)
directory with four sub-directories. All the source code files are
contained in the \textit{source} sub-directory. The \textit{examples}
sub-directory contains different test cases that can help the user
to edit new input files. The \textit{build} sub-directory stores a
UNIX \texttt{makefile} that assembles the executable versions of the
code both in serial and parallel version with different compilers.
Note that JETSPIN may be compiled on any UNIX platform. The \texttt{makefile}
should be copied (and eventually modified) into the \textit{source}
sub-directory, where the code is compiled and linked. A list of targets
for several common workstations and parallel computers can be used
by the command \textquotedbl{}\texttt{make target }\textquotedbl{},
where \texttt{target} is one of the options reported in Tab \ref{Tab:targets}.
On Windows system we advice the user to compile JETSPIN under the
command-line interface Cygwin \cite{racine2000cygwin}. Finally, the
binary executable file can be run in the \textit{execute} sub-directory.

\begin{table}[H]
\begin{centering}
\begin{tabular}{ll}
\hline 
\textbf{target:}  & \textbf{meaning:}\tabularnewline
\hline 
\hline 
gfortran  & compile in serial mode using the GFortran compiler.\tabularnewline
gfortran-mpi  & compile in parallel mode using the GFortran compiler and the Open
Mpi library.\tabularnewline
cygwin  & compile in serial mode using the GFortran compiler under the command-line\tabularnewline
 & interface Cygwin for Windows.\tabularnewline
cygwin-mpi  & compile in parallel mode using the GFortran compiler and the Open
Mpi library\tabularnewline
 & under the command-line interface Cygwin for Windows (note a precompiled\tabularnewline
 & package of the Open Mpi library is already available on Cygwin).\tabularnewline
intel  & compile in serial mode using the Intel compiler.\tabularnewline
intel-mpi  & compile in parallel mode using the Intel compiler and the Intel Mpi
library.\tabularnewline
intel-openmpi  & compile in parallel mode using the Intel compiler and the Open Mpi
library.\tabularnewline
help  & return the list of possible target choices\tabularnewline
\hline 
\end{tabular}
\par\end{centering}

\protect\caption{List of targets for several common workstations and parallel computers,
which can be used by the command \textquotedbl{}\texttt{make target}\textquotedbl{}.}

\label{Tab:targets} 
\end{table}

\section{Parallelization\label{sec:Parallelization}}

The parallel infrastructure of JETSPIN incorporates the necessary
data distribution and communication structures. The parallel strategy
underlying JETSPIN is the Replicated Data (RD) scheme\cite{smith1991molecular},
where fundamental data of the simulated system are reproduced on all
processing nodes. In simulations of nanofibers, the fundamental data
consist of position, velocity, and viscoelastic stress arrays of each
bead in which the nanojet is discretised (see below Cap \ref{chap:JETSPIN-Model-and}).
Further data defining mass and charge of each bead are also replicated.
However, all auxiliary data are distributed in equal portion of data
(as much as possible) for each processor. Despite being available
other parallel strategies such as the Domain Decomposition\cite{brown1993domain},
our experience has shown that such volume of data is by no means prohibitive
on current parallel computers.

By the RD scheme, we adopt in JETSPIN a simple communication strategy
of the communication between nodes, which is handled by global summation
routines. The module \texttt{version\_mod} located in the \texttt{parallel\_version\_mod.f90}
file contains all the global communication routines which exploit
the MPI (Message Passing Interface) library. Note that a FORTRAN90
compiler and an MPI implementation for the specific machine architecture
are required in order to compile JETSPIN in parallel mode. An alternative
version of the module \texttt{version\_mod} is located in the \texttt{serial\_version\_mod.f90}
file, and it can be easily selected by appropriate flags in the \texttt{makefile}
at compile time. By selecting this version, JETSPIN can also be run
on serial computers without modification, even though the code has
been designed to run on parallel computers.

The size system is strictly time-dependent as mentioned in Sec \ref{sec:Structure}
and, therefore, the memory of various service bookkeeping arrays is
dynamically distributed over all the processing nodes. In particular,
the bookkeeping array size, declared as \texttt{mxchunk}, is managed
by the routine \texttt{set\_mxchunk}. All the beads of the discretised
nanofiber are assigned at every time step to a specific node by the
routine \texttt{set\_chunk}, and their temporary data are stored in
bookkeeping arrays belonging to the assigned node. It is worth stressing
that the communication latency makes the parallelization efficiency
strictly dependent on the system size. Therefore, we only advice the
use of JETSPIN in parallel mode whenever the user expects a system
size with at least 50 beads for each node (further details in Ref
\cite{lauricella2015jetspin}).

\newpage{}

\chapter{JETSPIN Model and Algorithms\label{chap:JETSPIN-Model-and}}

\section*{Scope of Chapter}

This chapter describes the model and simulation algorithms incorporated
into JETSPIN.

\section{Equations of Motion\label{sec:Equation-of-Motion}}

The model implemented in JETSPIN is an extension of the Lagrangian
discrete model introduced by Reneker et al. \cite{reneker2000bending}
(in the following text referred as Ren model). The model provides
a compromise of efficiency and accuracy by representing the filament
as a series of $n$ beads (jet beads) at mutual distance $l$ connected
by viscoelastic elements. The length $l$ is typically larger than
the cross-sectional radius of the filament, but smaller than the characteristic
lengths of other observables of interest (e.g. curvature radius).
Each $i-th$ bead has mass $m_{i}$ and charge $q_{i}$ (not necessarily
equal for all the beads). The nanofiber is modelled as a viscoelastic
Maxwell fluid, so that the stress $\sigma_{i}$ for the element connecting
bead $i$ with bead $i+1$ is given by the viscoelastic constitutive
equation:

\begin{equation}
\frac{d\sigma_{i}}{dt}=\frac{G}{l_{i}}\frac{dl_{i}}{dt}-\frac{G}{\mu}\sigma_{i},\label{eq:stress-ode}
\end{equation}
where $l_{i}$ is the length of the element connecting bead $i$ with
bead $i+1$, $G$ is the elastic modulus, $\mu$ the viscosity of
the fluid jet, and $t$ the time. Given $a_{i}$ the cross-sectional
radius of the filament at the bead $i$, the viscoleastic force $\vec{\textbf{f}}_{\upsilon e}$
pulling bead $i$ back to $i-1$ and towards $i+1$ is

\begin{equation}
\vec{\textbf{f}}_{\upsilon e,i}=-\pi a_{i}^{2}\sigma_{i}\cdot\vec{\textbf{t}}_{i}+\pi a_{i+1}^{2}\sigma_{i+1}\cdot\vec{\textbf{t}}_{i+1}\text{,}
\end{equation}
where $\vec{\textbf{t}}_{i}$ is the unit vector pointing bead $i$
from bead $i-1$. The surface tension force $\vec{\textbf{f}}_{st}$
for the $i-th$ bead is given by

\begin{equation}
\vec{\textbf{f}}_{st,i}=k_{i}\cdot\pi\left(\frac{a_{i}+a_{i-1}}{2}\right)^{2}\alpha\cdot\vec{\textbf{c}}_{i}\text{,}
\end{equation}
where $\alpha$ is the surface tension coefficient, $k_{i}$ is the
local curvature, and $\vec{\textbf{c}}_{i}$ is the unit vector pointing
the center of the local curvature from bead $i$. Note the force $\vec{\textbf{f}}_{st}$
is acting to restore the rectilinear shape of the bending part of
the jet.

In the electrospinning experimental configuration an intense external
electric potential $\Phi_{0}$ is applied between the spinneret and
a conducting collector located at distance $h$ from the injection
point. As consequence, each $i-th$ bead endures the electric force

\begin{equation}
\vec{\textbf{f}}_{el,i}=e_{i}\frac{\Phi_{0}}{h}\cdot\vec{\textbf{x}}\text{,}
\end{equation}
where $\vec{\textbf{x}}$ is the unit vector pointing the collector
from the spinneret and assumed along the $x$ axis. Note that in Ren
model the electric field $\Phi$ is assumed to be static in order
to avoid the computationally expensive integration of Poisson equation,
whereas in reality $\Phi$ is depending on the net charge of the nanofiber
so as to maintain constant the potential at the electrodes. The latter
issue was elegantly addressed by Kowalewski et al. \cite{kowalewski2009modeling},
and its implementation in JETSPIN will be planned.

The net Coulomb force $\vec{\textbf{f}}_{c}$ acting on the $i-th$
bead from all the other beads is given by

\begin{equation}
\vec{\textbf{f}}_{c,i}=\sum_{\substack{j=1\\
j\neq i
}
}^{n}\frac{q_{i}q_{j}}{R_{ij}^{2}}\cdot\vec{\textbf{u}}_{ij}\text{,}
\end{equation}
where $R_{ij}=\left[\left(x_{i}-x_{j}\right)^{2}+\left(y_{i}-y_{j}\right)^{2}+\left(z_{i}-z_{j}\right)^{2}\right]^{1/2}$,
and $\vec{\textbf{u}}_{ij}$ is the unit vector pointing the $i-th$
bead from $j-th$ bead.. Although the Ren model provides a reasonable
description for the spiral motion of the jet, the last term $\vec{\textbf{f}}_{c}$
introduces mathematical inconsistencies due to the discretization
of the fiber into point-charges. Indeed, the charge induces a field
on the outer shell of the fiber and not on the center line (as in
the implemented model). Different approaches were developed to overcome
this issue which usually imply strong approximations\cite{feng2002stretching,feng2003stretching,hohman2001electrospinning}.
Other strategies use a less crude approximation by accounting for
the actual electrostatic form factors between two interacting sections
of a charged fiber\cite{kowalewski2005experiments} or involving more
sophisticated methods which exploit the tree-code hierarchical force
calculation algorithm\cite{kowalewski2009modeling}. The implementation
in JETSPIN of methods based on tree-code hierarchical force calculation
algorithm will be considered in future.

Although usually much smaller than the other driving forces, the body
force due to the gravity is computed in the model by the usual expression

\begin{equation}
\vec{\textbf{f}}_{g,i}=m_{i}g\cdot\vec{\textbf{x}},
\end{equation}
where $g$ is the gravitational acceleration.

As shown in experimental evidences \cite{spinning1991science}, the
air drag force affects the dynamics of the nanofiber. The extended
stochastic model developed in Refs \cite{lauricella2014electrospinning,lauricella2015langevin}
was included in JETSPIN to reproduce the aerodynamic effects. In particular,
the air drag is modelled by adding two force terms: a random term
and a dissipative term. The dissipative air drag term is usually dependent
on the geometry of the jet, which changes in time, and it combines
longitudinal and lateral components. Based on experimental results\cite{spinning1991science},
the longitudinal component of the air drag dissipative force term
acting on bead $i$ is

\begin{equation}
\vec{\textbf{f}}_{air,i}=-0.65\pi a_{i}l_{i}\rho_{a}\left(2a_{i}/\nu_{a}\right)^{-0.81}\upsilon_{tan,i}^{1+0.19}\cdot\vec{\textbf{t}}_{i-1}
\end{equation}
where $\rho_{a}$ and $\nu_{a}$ are the air density and kinematic
viscosity, respectively, $\vec{\textbf{t}}_{i-1}$ is the unit vector
pointing bead $i$ from bead $i-1$, and $\upsilon_{tan,i}=\vec{\boldsymbol{\upsilon}_{i}}\cdot\vec{\textbf{t}}_{i-1}$
is the projection of the vector velocity of bead $i$ along the unit
vector $\vec{\textbf{t}}_{i-1}$. Further, it is possible to write
$\vec{\textbf{f}}_{air,i}$ as

\begin{equation}
\vec{\textbf{f}}_{air,i}=-m_{i}\gamma_{i}\,l_{i}^{0.905}\upsilon_{t,i}^{1+0.19}\cdot\vec{\textbf{t}}_{i-1}
\end{equation}
if we introduced the dissipative coefficient $\gamma_{i}$ equal to

\begin{equation}
\gamma_{i}=0.65\pi^{0.905}\frac{\rho_{a}}{m_{i}}\left(\frac{2}{\nu_{air}}\right)^{-0.81}V_{i}^{0.095},
\end{equation}
where $V_{i}=\pi a_{i}^{2}\,l_{i}$ is the jet volume represented
by the $i-th$ bead.

Following the expression introduced by A. Yarin\cite{yarin1993free,yarin2014fundamentals},
under a high-speed air drag the lateral component $\vec{\textbf{f}}_{lift,i}$
of the aerodynamic dissipative force related to the flow speed is
given in the linear approximation (for small bending perturbations)
by 
\begin{equation}
\vec{\textbf{f}}_{lift,i}=-k_{i}\cdot\rho_{a}\upsilon_{i}^{2}\pi\left(\frac{a_{i}+a_{i-1}}{2}\right)^{2}\cdot\vec{\textbf{c}}_{i},
\end{equation}
where $k_{i}$ is still the local curvature, and $\vec{\textbf{c}}_{i}$
is the unit vector pointing the center of the local curvature from
bead $i$. The combined action of such longitudinal and lateral components
provide the dissipative force term acting on the $i-th$ bead

\begin{equation}
\vec{\textbf{f}}_{diss,i}=\vec{\textbf{f}}_{air,i}+\vec{\textbf{f}}_{lift,i}\:.
\end{equation}
Instead, the random force term for the $i-th$ bead has the form

\begin{equation}
\vec{\textbf{f}}_{rand,i}=\sqrt{2m_{i}^{2}D_{\upsilon}}\cdot\vec{\boldsymbol{\eta}_{i}}(t),
\end{equation}
where $D_{\upsilon}$ denotes a generic diffusion coefficient in velocity
space (which is assumed constant and equal for all the beads), and
$\vec{\boldsymbol{\eta}_{i}}$ is a three-dimensional vector, whereof
each component $\eta$ is an independent stochastic process that is
nowhere differentiable with $<\eta\left(t_{1}\right)\eta\left(t_{2}\right)>=\delta\left(\left|t_{2}-t_{1}\right|\right)$,
and $<\eta\left(t\right)>=0$. Note that for the sake of simplicity
we assume $\eta=d\omega(t)/dt$, where $\omega(t)$ is a Wiener process.

The combined action of these forces governs the elongation of the
jet according to the Newton's equation providing a non-linear Langevin-like
stochastic differential equation:

\begin{equation}
m_{i}\frac{d\vec{\boldsymbol{\upsilon}}_{i}}{dt}=\vec{\textbf{f}}_{el,i}+\vec{\textbf{f}}_{c,i}+\vec{\textbf{f}}_{\upsilon e,i}+\vec{\textbf{f}}_{st,i}+\vec{\textbf{f}}_{g,i}+\vec{\textbf{f}}_{diss,i}+\vec{\textbf{f}}_{rand}\:\text{,}\label{eq:force-EOM}
\end{equation}
where $\vec{\boldsymbol{\upsilon}}_{i}$ is the velocity of the $i-th$
bead. The velocity $\vec{\boldsymbol{\upsilon}}_{i}$ satisfies the
kinematic relation:

\begin{equation}
\frac{d\vec{\textbf{r}}_{i}}{dt}=\vec{\boldsymbol{\upsilon}}_{i}\label{eq:pos-EOM}
\end{equation}
where $\vec{\textbf{r}}_{i}$ is the position vector of the $i-th$
bead, $\vec{\textbf{r}}_{i}=x_{i}\vec{\textbf{x}}+y_{i}\vec{\textbf{y}}+z_{i}\vec{\textbf{z}}$.
The three Eqs \ref{eq:stress-ode}, \ref{eq:force-EOM} and \ref{eq:pos-EOM}
form the set of equations of motion (EOM) governing the time evolution
of system. It is worth stressing that Eq \ref{eq:force-EOM} recovers
a deterministic EOM by imposing $\rho_{a}=0$ and $D_{\upsilon}=0$.

\section{Perturbations at the nozzle}

The spinneret nozzle is represented by a single mass-less point of
charge $\bar{q}$ fixed at $x=0$ (nozzle bead). Its charge $\bar{q}$
is assumed equal to the mean charge value of the jet beads. Such charged
point can be also interpreted as a small portion of jet which is fixed
to the nozzle. In JETSPIN it is possible to add small perturbations
to the $y_{n}$ and $z_{n}$ coordinates of the nozzle bead in order
to model fast mechanical oscillations of the spinneret\cite{coluzza2014ultrathin}.
Given the initial position of the nozzle

\begin{subequations}

\begin{equation}
y_{n}=A\cdot\cos\left(\varphi\right)
\end{equation}

\begin{equation}
z_{n}=A\cdot\sin\left(\varphi\right),
\end{equation}

\end{subequations}

the equations of motion for the nozzle bead are

\begin{subequations}

\begin{equation}
\frac{dy_{n}}{dt}=-\omega\cdot z_{n}
\end{equation}

\begin{equation}
\frac{dz_{n}}{dt}=\omega\cdot y_{n},
\end{equation}

\end{subequations}

where $A$ denotes the amplitude of the perturbation, while $\omega$
and $\varphi$ are its frequency and initial phase, respectively.

\section{Jet insertion\label{sub:Jet-insertion}}

The jet insertion at the nozzle is modelled as follows. For the sake
of simplicity, let us consider a simulation which starts with only
two bodies: a single mass-less point fixed at $x=0$ representing
the spinneret nozzle, and a bead modelling an initial jet segment
of mass $m_{i}$ and charge $e_{i}$ located at distance $l_{step}$
from the nozzle along the $x$ axis. Here, $l_{step}$ denotes the
length step used to discretise the jet in a sequence of beads. The
starting jet bead is assumed to have a cross-sectional radius $a_{0}$,
defined as the radius of the filament at the nozzle before the stretching
process. Applying the condition that the volume of the jet is conserved,
the relation $\pi a_{i}^{2}\,l_{i}=\pi a_{0}^{2}l_{step}$ is valid
for any $i-th$ bead. Furthermore, the starting jet bead has an initial
velocity $\upsilon_{s}$ along the $x$ axis equal to the bulk fluid
velocity in the syringe needle. Once this traveling jet bead is a
distance of $2\cdot l_{step}$ away from the nozzle, a new jet bead
(third body) is placed at distance $l_{step}$ from the nozzle along
the straight line joining the two previous bodies. Let us now label
$i-1$ the farthest bead from the nozzle, and $i$ the last inserted
bead. The $i-th$ bead is inserted with the initial velocity $\upsilon_{i}=\upsilon_{s}+\upsilon_{d}$,
where $\upsilon_{d}$ denotes the dragging velocity computed as

\begin{equation}
\upsilon_{d}=\frac{\upsilon_{i-i}-\upsilon_{s}}{2}.
\end{equation}

Here, the dragging velocity should be interpreted as an extra term
which accounts for the drag effect of the electrospun jet on the last
inserted segment. Note that the actual dragging velocity definition
was chosen in order to not alter the strain velocity term $\left(1/l_{i-1}\right)\cdot\left(dl_{i-1}/dt\right)$
of Eq \ref{eq:stress-ode} before and after the bead insertion.

\section{Dynamic refinement\label{sec:Dynamic-refinement}}

In the original Ren model, each bead represent a constant quantity
$V_{c}$ of jet volume, so that the relationship $\pi a_{i}^{2}\,l_{i}=\pi a_{0}^{2}l_{step}=V_{c}$
is valid for any $i-th$ bead. Despite such relationship can be sometimes
a useful advantage (as example, see Sec \ref{sub:Introduction-of-dimensionless}),
the radius reduction $a_{i}$ of the electrospun nanofiber provides
a significant increase of the discretization length $l_{i}$ value
along the jet path (also greater than two order of magnitude). As
consequence, it was observed in literature \cite{kowalewski2009modeling}
that the jet discretization close the collector becomes rather coarse
to model efficiently the nanofiber. To get around this problem a dynamic
refinement can be used in JETSPIN in order to maintain the the length
of the elements $l_{i}$ below a prescribed characteristic length
max $l_{max}$. Whenever a bead length $l_{i}$ is larger than $l_{max}$,
the nanofiber is refined by discretizing uniformly the jet at the
length step $l_{step}$ value given in input file. The discretization
is performed by Akima spline interpolations \cite{akima1970new,akima1991method}
of the main quantities describing the jet beads (positions, fiber
radius, stress, velocities). In order to perform the interpolation
we proceed as follows: First of all, we introduce the variable $\lambda\in[0,1]$
to parametrize the jet path, where $\lambda=0$ identifies the nozzle,
and $\lambda=1$ the jet at the collector. For each bead we compute
its respective $\lambda_{i}$ value by using the formula

\begin{equation}
\lambda_{i}=\frac{1}{l_{path}}\sum_{k=1}^{i}l_{k}
\end{equation}
where $l_{path}=\sum_{k=1}^{N_{beads}}l_{k}$ is the jet path length
from the nozzle to the collector, and $i=1$ and $i=N_{beads}$ denote
the jet bead at the nozzle and collector, respectively ($i=0$ denotes
the nozzle). The set of ${\lambda_{i}}$ values represent the mesh
used to build the cubic spline. Given a generic quantity $y$, the
data $y_{i}=y\left(\lambda_{i}\right)$ are tabulated and used in
order to compute the coefficients of the \textit{Akima spline}, following
the algorithm reported in Ref \cite{akima1991method}. Then, we enforce
a uniform parametrization of the nanofiber by imposing all the lengths
of the elements equal to $l_{step}$. The new mesh ${\lambda_{i}^{*}}$
is defined as

\begin{equation}
\lambda_{0}^{*}=0\;\lambda_{i}^{*}=\frac{i}{N_{beads}^{*}},\;i=1,2,\ldots,N_{beads}^{*}\text{.}
\end{equation}
where $N_{beads}^{*}=l_{path}/l_{step}$ represents the number of
jet beads in the new representation. Finally, the new values $y\left(\lambda_{i}^{*}\right)$
are computed for any $i-th$ bead by the \textit{spline interpolation}.
The procedure is repeated for positions, fiber radius, stress and
velocities of the jet beads in order to provide the respective quantities
in the new mesh ${\lambda_{i}^{*}}$. It is worth to sressing that
the volume $V_{i}=\pi a_{i}^{2}\,l_{i}$ represented by each bead
is not anymore constant, and, consequently, $\pi a_{i}^{2}\,l_{i}\neq\pi a_{0}^{2}l_{step}$.
In the following text we will refer to this emended version of the
Ren model as DyRen model.

\section{Dimensionless quantities\label{sub:Introduction-of-dimensionless}}

In JETSPIN all the variables are automatically rescaled and stored
in dimensionless units. In order to adopt a dimensionless form of
the equations of motion, we use the dimensionless scaling procdure
proposed by Reneker et al.\cite{reneker2000bending}. We define a
characteristic length

\begin{equation}
L_{0}=\sqrt{\frac{\bar{q}^{2}}{\pi a_{0}^{2}G}}=l_{step}\sqrt{\frac{\pi a_{0}^{2}\rho_{V}^{2}}{G}},\label{eq:Length-scale}
\end{equation}
where we write the charge $q$ as $\pi a_{0}^{2}l_{step}\rho_{V}$,
denoting $\rho_{V}$ the electric volume charge density of the filament.
Further, we divide the time $t$ and the stress $\sigma$ by their
respective characteristic scales reported in Tab\ref{tab:tabella-adimensionale}.
Applying the condition that the volume of the jet is conserved $\pi a_{i}^{2}\,l_{i}=V_{i}$
for any $i-th$ bead, we rewrite the EOM as:

\begin{subequations}

\begin{equation}
\frac{d\vec{\bar{\textbf{r}}}_{i}}{d\bar{t}}=\vec{\boldsymbol{\bar{\upsilon}}}_{\,i}\label{eq:EOM-A1}
\end{equation}

\begin{equation}
\frac{d\bar{\sigma}_{i}}{d\bar{t}}=\frac{1}{\bar{l}_{i}}\frac{d\bar{l}_{i}}{d\bar{t}}-\bar{\sigma}_{i}\label{eq:EOM-A2}
\end{equation}

\begin{equation}
\begin{alignedat}{1}\frac{d\vec{\boldsymbol{\bar{\upsilon}}}_{\,i}}{d\bar{t}}= & \Phi\cdot\vec{\textbf{x}}+\sum_{\substack{j=1\\
j\neq i
}
}^{n}\frac{Q_{ij}}{\bar{R}_{ij}^{2}}\cdot\vec{\textbf{u}}_{ij}-F_{ve,i}\bar{V}_{i}\frac{\bar{\sigma}_{i}}{\bar{l}_{i}}\cdot\vec{\textbf{t}}_{i}+F_{ve,i+1}\bar{V}_{i+1}\frac{\bar{\sigma}_{i+1}}{\bar{l}_{i+1}}\cdot\vec{\textbf{t}}_{i+1}\\
 & +A_{i}\frac{\bar{k_{i}}}{4}\left(\frac{\sqrt{\bar{V}_{i}}}{\sqrt{\bar{l}_{i}}}+\frac{\sqrt{\bar{V}_{i-1}}}{\sqrt{\bar{l}_{i-1}}}\right)^{2}\cdot\vec{\textbf{c}}_{i}+F_{g}\cdot\vec{\textbf{x}}-\Gamma_{i}\,\bar{l}_{i}^{0.905}\bar{\upsilon}_{tan,i}^{1+0.19}\cdot\vec{\textbf{t}}_{i-1}-\\
 & -\Lambda_{i}\frac{\bar{k_{i}}\bar{\upsilon}_{tan,i}^{2}}{4}\left(\frac{\sqrt{\bar{V}_{i}}}{\sqrt{\bar{l}_{i}}}+\frac{\sqrt{\bar{V}_{i-1}}}{\sqrt{\bar{l}_{i-1}}}\right)^{2}\cdot\vec{\textbf{c}}_{i}+\sqrt{2\Theta_{\upsilon}}\cdot\vec{\boldsymbol{\eta}_{i}}(\bar{t})
\end{alignedat}
\label{eq:EOM-A3}
\end{equation}

\end{subequations}

where we used the dimensionless derived variables and groups defined
in Tab\ref{tab:tabella-adimensionale}. It is worth stressing that
the viscoelastic and surface tension force terms are slightly different
from the dimensionless form provided by Reneker et al. \cite{reneker2000bending},
since we are considering the DyRen model presented in Sec \ref{sec:Dynamic-refinement}.
In particular, we do not use the relation $\pi a_{i}^{2}\,l_{i}=\pi a_{0}^{2}l_{step}$
since it is valid only for the Ren model (see discussion in Sec \ref{sec:Dynamic-refinement}).

Finally, the dimensionless EOM of the nozzle are given as

\begin{subequations}

\begin{equation}
\frac{d\bar{y}_{i}}{d\bar{t}}=-K_{s}\cdot\bar{z}_{i}
\end{equation}

\begin{equation}
\frac{d\bar{z}_{i}}{d\bar{t}}=K_{s}\cdot\bar{y}_{i}\:,
\end{equation}

\end{subequations}

with the dimensionless parameter $K_{s}$ defined in Tab\ref{tab:tabella-adimensionale}.

\begin{table}[H]
\begin{centering}
\begin{tabular}{cc}
\hline 
\multicolumn{2}{c}{Characteristic Scales}\tabularnewline
\hline 
\hline 
$L_{0}=l_{step}\sqrt{\frac{\pi a_{0}^{2}\rho_{V}^{2}}{G}}$  & $t_{0}=\frac{\mu}{G}$\tabularnewline
$\sigma_{0}=G$  & \tabularnewline
\hline 
\multicolumn{2}{c}{Dimensionless Quantities}\tabularnewline
\hline 
\hline 
$\bar{l}_{i}={\displaystyle \frac{l_{i}}{L_{0}}}$  & $\bar{R}_{ij}={\displaystyle \frac{R_{ij}}{L_{0}}}$\tabularnewline
$\bar{\sigma}_{i}={\displaystyle \frac{\sigma_{i}}{\sigma_{0}}}$  & $\bar{\upsilon}_{i}=\upsilon_{i}{\displaystyle \frac{t_{0}}{L_{0}}}$ \tabularnewline
$\bar{k_{i}}=k_{i}L_{0}$  & $\bar{\upsilon}_{tan,i}=\upsilon_{tan,i}{\displaystyle \frac{t_{0}}{L_{0}}}$ \tabularnewline
$\bar{V}_{i}={\displaystyle \frac{V_{i}}{L_{0}^{3}}}$  & \tabularnewline
\hline 
\multicolumn{2}{c}{Dimensionless Groups}\tabularnewline
\hline 
\hline 
$\Phi_{i}={\displaystyle \frac{q_{i}\Phi_{0}\mu^{2}}{m_{i}h\,L_{0}\,G^{2}}}$  & $Q_{ij}={\displaystyle \frac{q_{i}q_{j}\mu^{2}}{m_{i}L_{0}^{3}G^{2}}}$\tabularnewline
$F_{ve,i}={\displaystyle \frac{L_{0}\mu^{2}}{m_{i}G}}$  & $A_{i}={\displaystyle \frac{\alpha\mu^{2}}{m_{i}G^{2}}}$\tabularnewline
$F_{g}={\displaystyle \frac{g\mu^{2}}{L_{0}G^{2}}}$  & $K_{s}=\omega{\displaystyle \frac{\mu}{G}}$\tabularnewline
$H={\displaystyle \frac{h}{L_{0}}}$  & $L_{step}={\displaystyle \frac{l_{step}}{L_{0}}}$\tabularnewline
$\Gamma_{i}={\displaystyle \gamma_{i}\cdot t_{0}L_{0}^{0.905}\left(\frac{L_{0}}{t_{0}}\right)^{0.19}}$  & $\Theta_{\upsilon}=D_{\upsilon}\cdot{\displaystyle \frac{t_{0}^{3}}{L_{0}^{2}}}$\tabularnewline
$\Lambda_{i}={\displaystyle \frac{\rho_{a}L_{0}^{2}}{m_{i}}}$  & \tabularnewline
\hline 
\end{tabular}
\par\end{centering}

\protect\protect\protect\caption{Definitions of the characteristic scales, dimensionless derived variables,
and groups employed in the text.}

\label{tab:tabella-adimensionale} 
\end{table}

\section{Integration schemes}

In order to integrate the homogeneous differential equations of motion
we discretise time as $t_{i}=t_{0}+i\Delta t$ with $i=1,\ldots,n_{steps}$,
where $n_{steps}$ denotes the number of sub-intervals.  Note that
in JETSPIN the time step $\Delta t$ is automatically rescaled by
the quantity $\tau$, in accordance with the mentioned dimensionless
scaling convention.
Three different integration schemes are available for deterministic simulations: the first-order accurate Euler
scheme, the second-order accurate Heun scheme (sometimes called second-order
accurate Runge-Kutta), and the fourth-order accurate Runge-Kutta scheme\cite{press2007numerical}.
The user can select a specific scheme by using appropriate keys in
the input file, as described in Sec \ref{sec:Input-file}.
For the integration of stochastic simulations the explicit strong order scheme by Platen \cite{platen1987derivative,kloeden1992numerical,platen2010numerical} was implemented,
whereof the order of strong convergence was evaluated
equal to 1.5. 
This scheme avoids the use of derivatives by corresponding finite
differences in the same way as Runge-Kutta schemes do for ODEs in
a deterministic setting, and it is briefly summarized as follows. 

Let us consider a Brownian motion vector process $\textbf{X}=\left\{ \textbf{X}_{t},t\right\} $ of
\textit{d}-dimensional satisfying the stochastic differential equation

\begin{equation}
\frac{d\textbf{X}}{dt}=\textbf{a}\left(t,X^{1},\ldots,X^{d}\right)+\textbf{b}d\Omega\label{eq:generic-sde}
\end{equation}
where $\textbf{a}$ and $\textbf{b}$ are vectors of \textit{d}-dimensional
usually called drift and diffusion vector coefficients, and $\Omega\left(t\right)$
denotes a Wiener process. Denoted $Y_{t}^{k}$ the approximation for
the \textit{k}-th component of the vector $\textbf{X}$ at time $t$, 
the integrator has the following form:

\begin{equation}
Y_{t+\varDelta t}^{k} = Y_{t}^{k}+b^{k}\Delta\Omega+
\frac{1}{2\sqrt{\Delta t}}\left[a^{k}\left(\widetilde{\boldsymbol{\Upsilon}}_{+}\right)-a^{k}\left(\widetilde{\boldsymbol{\Upsilon}}_{-}\right)\right]\Delta\varPsi+ 
 \frac{1}{4}\left[a^{k}\left(\widetilde{\boldsymbol{\Upsilon}}_{+}\right)+2a^{k}+a^{k}\left(\widetilde{\boldsymbol{\Upsilon}}_{-}\right)\right]\Delta t,
\label{eq:integrator-platen}
\end{equation}

with the vector supporting values

\begin{equation} 
\begin{alignedat}{1}\widetilde{\boldsymbol{\Upsilon}}_{\pm} & =\textbf{Y}_{t}+\textbf{a}\Delta t\pm\textbf{b}\sqrt{\Delta t}\\
 \widetilde{\boldsymbol{\Phi}}_{\pm} & =\widetilde{\boldsymbol{\Upsilon}}_{+}\pm\textbf{b}\left(\widetilde{\boldsymbol{\Upsilon}}_{+}\right)\sqrt{\Delta t}.
\end{alignedat}
\end{equation}

Here, $\Delta\Omega$ and $\Delta\varPsi$ indicate normally distributed
random variables constructed from two independent $N\left(0,1\right)$
standard Gaussian distributed random variables ($U_{1}$, $U_{2}$)
by means of the following linear transformation:

\begin{equation} 
\begin{alignedat}{1}\Delta\Omega & =U_{1}\sqrt{\Delta t}\\
 \Delta\varPsi & =\frac{1}{2}\Delta t^{3/2}\left(U_{1}+\frac{1}{\sqrt{3}}U_{2}\right).
\end{alignedat}
\label{eq:random-var-1}
\end{equation}

\newpage{}

\chapter{JETSPIN Data Files\label{chap:JETSPIN-Data-Files}}

\section*{Scope of Chapter}

This chapter describes all the input and output files of JETSPIN.

\section{Input file\label{sec:Input-file}}

In order to run JETSPIN simulations an input file has to be prepared,
which is a free-format and case-insensitive. The input file has to
be named \texttt{input.dat}, and it contains the selection of the
model system, integration scheme directives, specification of various
parameters for the model, and output directives. The input file does
not requires a specific order of key directives, and it is read by
the input parsing module. Every line is treated as a command sentence
(record). Records beginning with the symbol \# (commented) and blank
lines are not processed, and may be added to aid readability. Each
record is read in words (directives and additional keywords and numbers),
which are recognized as such by separation by one or more space characters.

As in the example given in \texttt{input test}, the last record is
a \texttt{finish} directive, which marks the end of the input data.
Before the \texttt{finish} directive, a wide list of directives may
be inserted (see Tab \ref{tab:inputlist}). The key \texttt{systype}
should be used to set the nanofiber model. In JETSPIN two models are
available: 1) the one dimensional model similar to the model of Cap
\ref{chap:JETSPIN-Model-and} but assuming the nanofiber to be straight
along the $\vec{\textbf{x}}$ axis, and, therefore, neglecting the
surface tension force $\vec{\textbf{f}}_{st}$; 2) the three dimensional
model described in Cap \ref{chap:JETSPIN-Model-and}. Internally these
options are handled by the integer variable \texttt{systype}, which
assumes the values explained in Tab \ref{sec:Input-file}. Further
details of the 1-D model can be found in Refs \cite{pontrelli2014electrospinning,carroll2011discretized}.
A series of variables are mandatory and have to be defined. For example,
\texttt{timestep}, \texttt{final time}, \texttt{initial length}, etc.
(see underlined directives in Tab \ref{tab:inputlist}). A missed
definition of any mandatory variable will call an error banner on
the terminal. The user can use two possible ways to define the initial
geometry of a nanofiber both given the mandatory directive \texttt{initial
length}: 1) the discretization step length of nanofiber is specified
by the directive \texttt{resolution}, which causes automatically the
setting of the jet segments number (the number of segment in which
the jet is discretised); 2) the jet segments number is declared by
the directive \texttt{points}, while the value of the discretization
step length is automatically set by the program (as in Example \textbf{Test
Case 3} in Sec \ref{sec:Test-Case-3:}). The directive \texttt{cutoff}
indicates the length of the upper and lower proximal jet sections,
which interact via Coulomb force on any bead. It is worth stressing
that the length value is set at the nozzle, so that the effective
cutoff increases along the simulation as the nanofiber is stretched.

The user should pay special attention in choosing the time integration
step given by the directive \texttt{timestep}, whose detailed considerations
are provided in Ref \cite{lauricella2015jetspin}. The reader is referred
to Tab \ref{tab:inputlist} for a complete listing of all directives
defining the electrospinning setup parameters. Not all these quantities
are mandatory, but the user is informed that whenever a quantity is
missed, it is usually assumed equal to zero by default (exceptions
are stressed in Tab \ref{tab:inputlist}). Note all the quantities
have to be expressed in centimeter\textendash gram\textendash second
unit system (e.g. charge in statcoulomb, electric potential in statvolt,
etc.).

\begin{table}[H]
\begin{centering}
\begin{tabular}{ll}
\hline 
\textbf{directive:}  & \textbf{meaning:}\tabularnewline
\hline 
\hline 
airdrag yes  & active the inclusion of the airdrag force in the model\tabularnewline
 & (note that if yes the directive ``integrator 4'' is mandatory)\tabularnewline
airdrag airdensity $f$  & mass density of the air surrounding the nanofiber (default: 0)\tabularnewline
airdrag airviscosity $f$  & kinematic viscosity of the air surrounding the nanofiber (default:
1)\tabularnewline
airdrag airvelocity $f$  & velocity of the air with respect to the \textit{z} axis (default:
0)\tabularnewline
airdrag amplitude $f$  & amplitude of the randomic force rescaled by $\gamma$ (default: 1)\tabularnewline
 & (note that the dissipation coefficient $\gamma$ is automatically\tabularnewline
 & computed by JETSPIN using the input data)\tabularnewline
\uline{collector distance} $f$  & distance of collector along the x axis from the nozzle\tabularnewline
 & \quad{}(note the nozzle is assumed in the origin)\tabularnewline
cutoff $f$  & length of the proximal jet sections interacting by Coulomb \tabularnewline
 & \quad{}force (default: equal to the collector distance)\tabularnewline
\uline{density mass} $f$  & mass density of the nanofiber\tabularnewline
\uline{density charge} $f$  & electric volume charge density of the nanofiber\tabularnewline
dynam refin yes & active the dynamic refinement during the simultion\tabularnewline
dynam refin threshold $f$  & set the threshold beyond which the refinement is applied\tabularnewline
dynam refin every $f$  & the refinement is applied every $f$ seconds\tabularnewline
dynam refin start $f$  & the refinement is applied only if the simulation time is greater \tabularnewline
 & than or equal to $f$ seconds\tabularnewline
\uline{elastic modulus} $f$  & elastic modulus of the nanofiber\tabularnewline
external potential $f$  & electric potential between the nozzle and collector\tabularnewline
\uline{final time} $f$  & set the end time of the simulation \tabularnewline
\uline{finish}  & close the input file (last data record)\tabularnewline
gravity yes  & active the inclusion of the gravity force in the model\tabularnewline
\uline{initial length} $f$  & length of the nanofiber at the initial time\tabularnewline
\uline{integrator} $i$  & set the integration scheme. The \textit{integer} can be '1' for\tabularnewline
 & \quad{}Euler, '2' for Heun, '3' for Runge-Kutta, and '4'\tabularnewline
 & \quad{}for the Platen scheme (only for airdrag force activated).\tabularnewline
inserting yes  & inject new beads as explained in Subsec \tabularnewline
\uline{nozzle cross} $f$  & cross section radius of the nanofiber at the nozzle\tabularnewline
nozzle stress $f$  & viscoelastic stress of the nanofiber at the nozzle\tabularnewline
nozzle velocity $f$  & bulk fluid velocity of the nanofiber at the nozzle\tabularnewline
perturb yes  & active the periodic perturbation at the nozzle\tabularnewline
perturb freq $f$  & frequency of the perturbation at the nozzle\tabularnewline
perturb ampl $f$  & amplitude of the perturbation at the nozzle\tabularnewline
points $i$  & number of segment in which the jet is discretised\tabularnewline
print list $s_{1}\,\ldots$  & print on terminal statistical data as indicated by symbolic\tabularnewline
 & \quad{}strings (see Tab \ref{tab:outputlist} for detailed informations)\tabularnewline
print time $f$  & print data on terminal and output file every $f$ seconds\tabularnewline
print xyz $f$  & print the trajectory in XYZ file format in centimeters\tabularnewline
 & \quad{}every $f$ seconds \tabularnewline
print xyz frame $f$  & print the jet geometry in a single XYZ file format in centimeters\tabularnewline
 & \quad{}every $f$ seconds (default name of files: frame\%06d.xyz)\tabularnewline
print xyz maxnum $i$  & print the XYZ file format with $i$ beads starting from the\tabularnewline
 & \quad{}nearest bead to the collector (default: 100)\tabularnewline
print xyz rescale $f$  & print the XYZ file format data rescaled by $f$\tabularnewline
printstat list $s_{1}\,\ldots$  & print on output file statistical data as indicated by symbolic\tabularnewline
 & \quad{}strings (see Tab \ref{tab:outputlist} for detailed informations)\tabularnewline
removing yes  & remove beads at collector\tabularnewline
resolution $f$  & discretization step length of nanofiber\tabularnewline
seed $i$  & seed control to the internal random number generator\tabularnewline
surface tension $f$  & surface tension coefficient of the nanofiber\tabularnewline
\uline{system} $i$  & set the nanofiber model. The \textit{integer} can be equal to \tabularnewline
 & \quad{}'1' for select the 1d-model or '3' for the 3d-model\tabularnewline
\uline{timestep} $f$  & set the time step for the integration scheme\tabularnewline
\uline{viscosity} $f$  & viscosity of the nanofiber\tabularnewline
\hline 
\end{tabular}
\par\end{centering}

\protect\protect\protect\caption{Here, we report the list of directives available in JETSPIN. Note
$i$, $f$, and $s$ denote an integer number, a floating point number,
and a string, respectively. The underlined directives are mandatory.
The default value is zero (exceptions are stressed in parentheses).}

\label{tab:inputlist} 
\end{table}

\section{Output files\label{sec:Output-files}}

A series of specific directives causes the writing of output files
(see Tab \ref{tab:inputlist}). In JETSPIN three output files can
be written: 1) the file \texttt{statdat.dat} containing time-dependent
statistical data of simulated process; 2) the file \texttt{traj.xyz}
reporting the nanofiber trajectory in XYZ format file; 3) the files
\texttt{frame\%06d.xyz} reporting the nanofiber geometries in splitted
XYZ format files (a single file for each frame).

Various statistical data can be written on the file \texttt{statdat.dat}
, and the user can select them in input using the directive \texttt{printstat
list} followed by appropriate symbolic strings, whose the list with
corresponding meanings is reported in Tab \ref{tab:outputlist}. The
selected observables will be printed at time interval on the same
line following the order specified in input file. In the same way,
a list of statistical data can be printed on computer terminal using
the directive \texttt{print list} followed by the symbolic strings
of Tab \ref{tab:outputlist}.

The file \texttt{traj.xyz} is written as a continuous series of XYZ
format frames taken at time interval, so that can be read by suitable
visualization programs (e.g. VMD-Visual Molecular Dynamics \cite{humphrey1996vmd},
UCSF Chimera \cite{pettersen2004ucsf}, etc.) to generate animations.
The number of elements contained in the file is kept constant equal
to the value specified by the directive \texttt{print xyz maxnum},
since few programs (e.g. VMD) do not manage a variable number of elements
along the simulations. If the actual number of beads is lower than
the given constant \texttt{maxnum}, the extra elements are printed
in the origin point.

\begin{table}[H]
\begin{centering}
\begin{tabular}{ll}
\hline 
\textbf{keys:}  & \textbf{meaning:}\tabularnewline
\hline 
\hline 
t  & unscaled time\tabularnewline
ts  & scaled time \tabularnewline
x, y, z  & unscaled coordinates of the farest bead \tabularnewline
xs, ys, zs  & scaled coordinates of the farest bead\tabularnewline
st  & unscaled stress of the farest bead\tabularnewline
sts  & scaled stress of the farest bead\tabularnewline
vx, vy, vz  & unscaled velocities of the farest bead \tabularnewline
vxs, vys, vzs  & scaled velocities of the farest bead \tabularnewline
yz  & unscaled normal distance from the x axis of the farest bead\tabularnewline
yzs  & scaled normal distance from the x axis of the farest bead\tabularnewline
mass  & last inserted mass at the nozzle \tabularnewline
q  & last inserted charge at the nozzle\tabularnewline
cpu  & time for every print interval\tabularnewline
cpur  & remaining time to the end \tabularnewline
cpue  & elapsed time \tabularnewline
n  & number of beads used to discretise the jet\tabularnewline
f  & index of the first bead \tabularnewline
l  & index of the last bead \tabularnewline
\uline{curn}  & current at the nozzle \tabularnewline
\uline{curc}  & current at the collector \tabularnewline
\uline{vn}  & velocity modulus of jet at the nozzle \tabularnewline
\uline{vc}  & velocity modulus of jet at the collector\tabularnewline
\uline{svc}  & strain velocity at the collector \tabularnewline
\uline{mfn}  & mass flux at the nozzle \tabularnewline
\uline{mfc}  & mass flux at the collector \tabularnewline
\uline{rc}  & radius of jet at the collector \tabularnewline
\uline{rrr}  & radius reduction ratio of jet \tabularnewline
\uline{lp}  & length path of jet \tabularnewline
\uline{rlp}  & length path of jet divided by the collector distance\tabularnewline
\hline 
\end{tabular}
\par\end{centering}

\protect\protect\caption{In JETSPIN a series of instantaneous and statistical data are available
to be printed by selecting the appropriate key. Here, the list of
symbolic string keys is reported with their corresponding meanings.
Note by \textit{scaled} we mean that the observable was rescaled by
the characteristic values provided in Tab\ref{tab:tabella-adimensionale}.
The underlined keys correspond to data which are averaged on the time
interval given by the directive \texttt{print time} in input file.
Note by \textit{farest bead} we mean the farest bead from the nozzle
(with greatest $x$ value). Note all the quantities are expressed
in centimeter\textendash gram\textendash second unit system, excepted
the dimensionless scaled observables.}

\label{tab:outputlist} 
\end{table}

\newpage{}

\chapter{Example Simulations}

\section*{Scope of Chapter}

This chapter describes the standard test cases for JETSPIN, the input
and output files for which are in the \textit{examples} sub-directory.

\section{Test Case 1: 1-D case}

This input file (located in the \textit{input-1} sub-directory) provides
the dimensionless parameter values $Q=12$, $V=2$ and $F_{ve}=12$,
which have been already used as reference case in Refs \cite{reneker2000bending,pontrelli2014electrospinning}.

\section{Test Case 2: 1-D case with bead insertion at the nozzle activated}

This input file (located in the \textit{input-2} sub-directory) provides
the dimensionless parameter values $Q=12$, $V=2$ and $F_{ve}=12$,
as in the previous test case, but here we have activated the injection
of new beads by the directive \texttt{inserting yes} in the input
file.

\section{Test Case 3: Three-D simulations of a process leading to polymer
nanofibers\label{sec:Test-Case-3:}}

The electrospinning of polyvinylpyrrolidone (PVP) nanofibers is a
prototypical process, which has been largely investigated in literature.\cite{yarin2014fundamentals,pisignanoelectrospinning,persano2013industrial}
By the input located in the \textit{input-3} sub-directory, we simulate
the electrospinning process of PVP solutions. Then, the theoretical
results predicted by the models are compared with the aforementioned
experimental data. In particular, we reproduce an experiment in which
a solution of PVP (molecular weight = 1300 kDa) is prepared by a mixture
of ethanol and water (17:3 v:v), at a concentration of about 2.5 wt\%.
The applied voltage is in a range around 10 kV, and the collector
is placed at distance 16 cm from the nozzle, which has radius $250\,$micron
(further details are provided in Ref. \cite{montinaro2015electrospinning,lauricella2015jetspin}
). As rheological properties of such system we consider the zero-shear
viscosity $\mu_{0}=0.2\,\text{g}/(\text{cm}\cdot\text{s})$ \cite{yuya2010morphology,buhler2005polyvinylpyrrolidone},
the elastic modulus $G=5\cdot10^{4}\,\text{g}/(\text{cm}\cdot\text{s}^{2})$
\cite{morozov2012water}, and the surface tension $\alpha=21.1\,\text{g}/\text{s}^{2}$
\cite{yuya2010morphology}. We use for the simulation a viscosity
value $\mu$ which is two order of magnitude larger than the zero-shear
viscosity $\mu_{0}$ reported, since the strong longitudinal flows
we are dealing with can lead to an increase of the extensional viscosity
from $\mu_{0}$, as already observed in literature.\cite{reneker2000bending,yarin1993free}
The mass density is equal to $0.84\,\text{g}/\text{cm}^{3}$, while
the charge density was estimated by experimental observations of the
current measured at the nozzle. For convenience, all the simulation
parameters are summarized in Tab\ref{tab:simulation-param}.

\begin{table}
\begin{centering}
\begin{tabular}{cccccccccc}
\hline 
$\rho$  & $\rho_{q}$  & $a_{0}$  & $\upsilon_{s}$  & $\alpha$  & $\mu$  & $G$  & $V_{0}$  & $\omega$  & $A$\tabularnewline
($\text{kg}/\text{m}^{3}$)  & ($\text{C}/\text{L}$)  & ($\text{cm}$)  & ($\text{cm/s}$)  & (mN/m)  & (Pa$\cdot$s)  & (Pa)  & (kV)  & ($\text{s}^{-1}$)  & ($\text{cm}$)\tabularnewline
\hline 
\hline 
840  & $2.8\cdot10^{-7}$  & $5\cdot10^{-3}$  & 0.28  & 21.1  & 2.0  & 50000  & 9.0  & $10^{4}$  & $10^{-3}$\tabularnewline
\hline 
\end{tabular}
\par\end{centering}

\protect\protect\caption{Simulation parameters for the simulation of PVP nanofibers. The headings
used are as follows: $\rho$: density, $\rho_{q}$: charge density,
$a_{0}$: fiber radius at the nozzle, $\upsilon_{s}$: bulk fluid
velocity in the syringe needle, $\alpha$: surface tension, $\mu$:
viscosity, $G$ : elastic modulus, $V_{0}$ : applied voltage bias,
$\omega$: frequency of perturbation, $A$ : amplitude of perturbation.
The bulk fluid velocity $\upsilon_{s}$ was estimated considering
that the solution was pumped at constant flow rate of 2 mL/h in a
needle of radius $250\,$micron.}

\label{tab:simulation-param} 
\end{table}

\newpage{}

\newpage{} 

\begin{thebibliography}{10}
\expandafter\ifx\csname url\endcsname\relax
  \def\url#1{\texttt{#1}}\fi
\expandafter\ifx\csname urlprefix\endcsname\relax\def\urlprefix{URL }\fi
\expandafter\ifx\csname href\endcsname\relax
  \def\href#1#2{#2} \def\path#1{#1}\fi

\bibitem{li2004electrospinning}
D.~Li, Y.~Wang, Y.~Xia, Electrospinning nanofibers as uniaxially aligned arrays
  and layer-by-layer stacked films, Advanced Materials 16~(4) (2004) 361--366.

\bibitem{greiner2007electrospinning}
A.~Greiner, J.~H. Wendorff, Electrospinning: a fascinating method for the
  preparation of ultrathin fibers, Angewandte Chemie International Edition
  46~(30) (2007) 5670--5703.

\bibitem{carroll2008nanofibers}
C.~P. Carroll, E.~Zhmayev, V.~Kalra, Y.~L. Joo, Nanofibers from electrically
  driven viscoelastic jets: modeling and experiments, Korea-Aust Rheol J 20
  (2008) 153--164.

\bibitem{huang2003review}
Z.-M. Huang, Y.-Z. Zhang, M.~Kotaki, S.~Ramakrishna, A review on polymer
  nanofibers by electrospinning and their applications in nanocomposites,
  Composites science and technology 63~(15) (2003) 2223--2253.

\bibitem{persano2013industrial}
L.~Persano, A.~Camposeo, C.~Tekmen, D.~Pisignano, Industrial upscaling of
  electrospinning and applications of polymer nanofibers: a review,
  Macromolecular Materials and Engineering 298~(5) (2013) 504--520.

\bibitem{yarin2014fundamentals}
A.~L. Yarin, B.~Pourdeyhimi, S.~Ramakrishna, Fundamentals and Applications of
  Micro and Nanofibers, Cambridge University Press, 2014.

\bibitem{wendorff2012electrospinning}
J.~H. Wendorff, S.~Agarwal, A.~Greiner, Electrospinning: materials, processing,
  and applications, John Wiley \& Sons, 2012.

\bibitem{pisignanoelectrospinning}
D.~Pisignano, Polymer Nanofibers: Building Blocks for Nanotechnology, Royal
  Society of Chemistry, 2013.

\bibitem{zeng2006numerical}
Y.~Zeng, Y.~Wu, Z.~Pei, C.~Yu, Numerical approach to electrospinning,
  International Journal of Nonlinear Sciences and Numerical Simulation 7~(4)
  (2006) 385--388.

\bibitem{feng2003stretching}
J.~Feng, Stretching of a straight electrically charged viscoelastic jet,
  Journal of Non-Newtonian Fluid Mechanics 116~(1) (2003) 55--70.

\bibitem{reneker2000bending}
D.~H. Reneker, A.~L. Yarin, H.~Fong, S.~Koombhongse, Bending instability of
  electrically charged liquid jets of polymer solutions in electrospinning,
  Journal of Applied physics 87~(9) (2000) 4531--4547.

\bibitem{yarin2001taylor}
A.~L. Yarin, S.~Koombhongse, D.~H. Reneker, Taylor cone and jetting from liquid
  droplets in electrospinning of nanofibers, Journal of Applied Physics 90~(9)
  (2001) 4836--4846.

\bibitem{fridrikh2003controlling}
S.~V. Fridrikh, H.~Y. Jian, M.~P. Brenner, G.~C. Rutledge, Controlling the
  fiber diameter during electrospinning, Physical review letters 90~(14) (2003)
  144502.

\bibitem{theron2004experimental}
S.~Theron, E.~Zussman, A.~Yarin, Experimental investigation of the governing
  parameters in the electrospinning of polymer solutions, Polymer 45~(6) (2004)
  2017--2030.

\bibitem{lu2006computer}
C.~Lu, P.~Chen, J.~Li, Y.~Zhang, Computer simulation of electrospinning. part
  i. effect of solvent in electrospinning, Polymer 47~(3) (2006) 915--921.

\bibitem{wang2013simulation}
X.~Wang, Y.~Liu, C.~Zhang, Y.~An, X.~He, W.~Yang, Simulation on electrical
  field distribution and fiber falls in melt electrospinning, Journal of
  nanoscience and nanotechnology 13~(7) (2013) 4680--4685.

\bibitem{coluzza2014ultrathin}
I.~Coluzza, D.~Pisignano, D.~Gentili, G.~Pontrelli, S.~Succi, Ultrathin fibers
  from electrospinning experiments under driven fast-oscillating perturbations,
  Physical Review Applied 2~(5) (2014) 054011.

\bibitem{racine2000cygwin}
J.~Racine, The cygwin tools: a gnu toolkit for windows, Journal of Applied
  Econometrics 15~(3) (2000) 331--341.

\bibitem{smith1991molecular}
W.~Smith, Molecular dynamics on hypercube parallel computers, Computer Physics
  Communications 62~(2) (1991) 229--248.

\bibitem{brown1993domain}
D.~Brown, J.~H. Clarke, M.~Okuda, T.~Yamazaki, A domain decomposition
  parallelization strategy for molecular dynamics simulations on distributed
  memory machines, Computer Physics Communications 74~(1) (1993) 67--80.

\bibitem{lauricella2015jetspin}
M.~Lauricella, G.~Pontrelli, I.~Coluzza, D.~Pisignano, S.~Succi, Jetspin: a
  specific-purpose open-source software for simulations of nanofiber
  electrospinning, submitted to Computer Physics Communications.

\bibitem{kowalewski2009modeling}
T.~A. Kowalewski, S.~Barral, T.~Kowalczyk, Modeling electrospinning of
  nanofibers, in: IUTAM symposium on modelling nanomaterials and nanosystems,
  Springer, 2009, pp. 279--292.

\bibitem{feng2002stretching}
J.~Feng, The stretching of an electrified non-newtonian jet: A model for
  electrospinning, Physics of Fluids (1994-present) 14~(11) (2002) 3912--3926.

\bibitem{hohman2001electrospinning}
M.~M. Hohman, M.~Shin, G.~Rutledge, M.~P. Brenner, Electrospinning and
  electrically forced jets. i. stability theory, Physics of Fluids
  (1994-present) 13~(8) (2001) 2201--2220.

\bibitem{kowalewski2005experiments}
T.~Kowalewski, S.~NSKI, S.~Barral, Experiments and modelling of electrospinning
  process, Technical Sciences 53~(4).

\bibitem{spinning1991science}
A.~Ziabicki, H.~Kawai, High-Speed Fiber Spinning: Science and Engineering
  Aspects, Krieger Publishing Co, 1991.

\bibitem{lauricella2014electrospinning}
M.~Lauricella, G.~Pontrelli, I.~Coluzza, D.~Pisignano, S.~Succi, Different
  regimes of the uniaxial elongation of electrically charged viscoelastic jets
  due to dissipative air drag, Submitted to Mech. Res. Comm., (2014).

\bibitem{lauricella2015langevin}
M.~Lauricella, G.~Pontrelli, D.~Pisignano, S.~Succi, Non-linear langevin model
  for the early-stage dynamics of electrospinning jets, Submitted to Molecular
  Physics, (2015).

\bibitem{yarin1993free}
A.~L. Yarin, Free liquid jets and films: hydrodynamics and rheology, Longman
  Scientific \& Technical Harlow, 1993.

\bibitem{akima1970new}
H.~Akima, A new method of interpolation and smooth curve fitting based on local
  procedures, Journal of the ACM (JACM) 17~(4) (1970) 589--602.

\bibitem{akima1991method}
H.~Akima, A method of univariate interpolation that has the accuracy of a
  third-degree polynomial, ACM Transactions on Mathematical Software (TOMS)
  17~(3) (1991) 341--366.

\bibitem{press2007numerical}
W.~H. Press, Numerical recipes 3rd edition: The art of scientific computing,
  Cambridge university press, 2007.

\bibitem{platen1987derivative}
E.~Platen, Derivative free numerical methods for stochastic differential
  equations, in: Stochastic Differential Systems, Springer, 1987, pp. 187--193.

\bibitem{kloeden1992numerical}
P.~E. Kloeden, E.~Platen, Numerical solution of stochastic differential
  equations, Vol.~23, Springer, 1992.

\bibitem{platen2010numerical}
E.~Platen, N.~Bruti-Liberati, Numerical solution of stochastic differential
  equations with jumps in finance, Vol.~64, Springer, 2010.

\bibitem{pontrelli2014electrospinning}
G.~Pontrelli, D.~Gentili, I.~Coluzza, D.~Pisignano, S.~Succi, Effects of
  non-linear rheology on the electrospinning process: a model study, Mechanics
  Research Communications 61 (2014) 41--46.

\bibitem{carroll2011discretized}
C.~P. Carroll, Y.~L. Joo, Discretized modeling of electrically driven
  viscoelastic jets in the initial stage of electrospinning, Journal of Applied
  Physics 109~(9) (2011) 094315.

\bibitem{humphrey1996vmd}
W.~Humphrey, A.~Dalke, K.~Schulten, Vmd: visual molecular dynamics, Journal of
  molecular graphics 14~(1) (1996) 33--38.

\bibitem{pettersen2004ucsf}
E.~F. Pettersen, T.~D. Goddard, C.~C. Huang, G.~S. Couch, D.~M. Greenblatt,
  E.~C. Meng, T.~E. Ferrin, Ucsf chimera: a visualization system for
  exploratory research and analysis, Journal of computational chemistry 25~(13)
  (2004) 1605--1612.

\bibitem{montinaro2015electrospinning}
M.~Montinaro, V.~Fasano, M.~Moffa, A.~Camposeo, L.~Persano, M.~Lauricella,
  S.~Succi, D.~Pisignano, Sub-ms dynamics of the instability onset of
  electrospinning, Submitted to Soft Matter.

\bibitem{yuya2010morphology}
N.~Yuya, W.~Kai, B.-S. Kim, I.~Kim, Morphology controlled electrospun
  poly(vinyl pyrrolidone) fibers: effects of organic solvent and relative
  humidity, Journal of Materials Science and Engineering with Advanced
  Technology.

\bibitem{buhler2005polyvinylpyrrolidone}
V.~B{\"u}hler, Polyvinylpyrrolidone excipients for pharmaceuticals: povidone,
  crospovidone and copovidone, Springer Science \& Business Media, 2005.

\bibitem{morozov2012water}
V.~N. Morozov, A.~Y. Mikheev, Water-soluble polyvinylpyrrolidone nanofilters
  manufactured by electrospray-neutralization technique, Journal of Membrane
  Science 403 (2012) 110--120.

\end{thebibliography}
\end{document}